\documentclass[10pt,a4paper,twocolumn]{article}

\usepackage[T1]{fontenc}
\usepackage[utf8]{inputenc}
\usepackage{lmodern}
\usepackage[margin=2cm,columnsep=0.7cm]{geometry}

\usepackage{amsmath}
\usepackage{amssymb}

\usepackage{graphicx}
\usepackage{booktabs}
\usepackage{makecell}

\usepackage[numbers,sort&compress]{natbib}
\usepackage{authblk}

\usepackage{xurl}
\usepackage[colorlinks=true,allcolors=blue]{hyperref}

\usepackage{xcolor}

\usepackage{abstract}
\begin{document}

\title{Automated Left Ventricle Strain Analysis from DENSE MRI}


\author[1,*]{Mohammad Naqizadeh Jahromi}
\author[2,3,4]{Ariel~J.~Hannum}
\author[2,3]{Zhan-Qiu Liu}
\author[2,3,4,5]{Daniel~B.~Ennis}
\author[1]{Dazhong Wu}
\author[1,6]{Luigi~E.~Perotti}

\affil[1]{Department of Mechanical and Aerospace Engineering, University of Central Florida, Orlando, FL 32816, USA}
\affil[2]{Department of Radiology, Stanford University, Stanford, CA 94305, USA}
\affil[3]{Division of Radiology, Veterans Administration Health Care System, Palo Alto, CA 94304, USA}
\affil[4]{Cardiovascular Institute, Stanford University, Stanford, CA 94305, USA}
\affil[5]{Department of Bioengineering, Stanford University, Stanford, CA 94305, USA}
\affil[6]{Biionix Cluster, University of Central Florida, Orlando, FL 32816, USA}
\affil[*]{\textit{Corresponding author:} Email: \href{mailto:mohammad.naqizadehjahromi@ucf.edu}{mohammad.naqizadehjahromi@ucf.edu}; ORCID: \href{https://orcid.org/0009-0008-7045-4848}{0009-0008-7045-4848}}

\renewcommand\Affilfont{\small}
\date{}


\twocolumn[
\begin{@twocolumnfalse}
\maketitle
\hrule
\vspace{0.5em}
\begin{onecolabstract} 
\noindent\textbf{Background:} Quantifying myocardial strains from Displacement ENcoding with Stimulated Echoes (DENSE) MRI provides insight into left ventricular performance, but current data processing pipelines remain limited in automation, robustness, and full left ventricle coverage. 

\noindent\textbf{Methods:} In this work, we present a fully automated framework for voxel-wise displacement and strain analysis of the left ventricle from DENSE MRI. Our method integrates a multi-channel nnU-Net model trained on both magnitude and phase images for segmentation, an automated phase unwrapping approach, an adaptation of the regularized spatiotemporal least squares (RSTLS) formulation for displacement computation, and strain calculations based on radial basis function interpolation. The proposed framework enables the estimation of displacements and strains across the left ventricle from a full stack of short-axis images. 

\noindent\textbf{Results:} Validation using a computational phantom shows the framework's overall resilience to data noise. Following computational phantom evaluation, the proposed framework is applied to analyze DENSE MRI short-axis slices acquired in forty-seven (N=47) healthy volunteers and covering the full left ventricle. Circumferential, longitudinal, and radial strains are presented throughout the cardiac cycle and at peak systole, showcasing the performance of the proposed framework. Results show that circumferential and longitudinal strains are computed accurately. However, challenges persist when estimating radial strain, especially in the subendocardial layer. A group-wise strain comparison is also performed based on volunteers' sex and age showing a significant effect of sex on circumferential and longitudinal strains.

\noindent\textbf{Conclusions:} Overall, the proposed framework provides an automated and modular pipeline to compute displacements and strains from DENSE MR images that can be scaled and generalized toward clinical applications.


\par\vspace{0.5em}
{\color{gray}\hrule}
\vspace{0.8em}
\end{onecolabstract}
\vspace{0.0em}
{\color{gray}\hrule}
\vspace{0.5em}

{\small

\vspace{1em}
}
\end{@twocolumnfalse}
]


\section{Introduction}

Cardiac strains can be used as markers for the early detection of cardiac abnormalities and dysfunction. Among available image acquisition techniques, Displacement ENcoding with Stimulated Echoes (DENSE) magnetic resonance imaging (MRI) stands out because it encodes tissue displacement directly into the image phase data, enabling high-resolution calculation of voxel-wise strains throughout the cardiac cycle~\cite{aletras1999dense}. Since computing strain measures requires the calculation of displacement spatial gradients, experimental noise present in the data may significantly distort strain maps. Robust displacement acquisition and subsequent conversion from Eulerian (encoded in the image phase data) to Lagrangian (required for strain calculations) descriptions are thus essential for accurate and clinically meaningful strain analysis.

Early advancements in DENSE image acquisition have primarily focused on enabling accurate and efficient displacement encoding under breath-hold conditions. Kim et al.~\cite{kim2004myocardial} introduced a breath-hold pulse sequence that combined \textit{complementary spatial modulation} for artifact suppression with \textit{echo-planar} imaging for rapid acquisition. To further enhance displacement fidelity, Epstein et al.~\cite{epstein2004displacement} adopted the Cosine ANd Sine modulation to ELiminate (CANSEL) method, which isolates the displacement-encoding echoes and suppresses artifact-generating signals. Building on these foundations, \textit{spiral} readout trajectories were introduced to improve signal-to-noise ratio and acquisition efficiency compared to \textit{echo-planar} imaging~\cite{zhong2010imaging}. These developments collectively ensured cleaner input data for downstream displacement and strain analysis.

Various post-processing methods, parallel to acquisition improvements, were developed to robustly extract displacement and strain data from the DENSE phase images. Spottiswoode et al.~\cite{spottiswoode2007tracking} introduced an approach that unwraps the phase data both spatially and temporally and fits (spatially and temporally) the displacement trajectories. After spatiotemporal phase unwrapping and trajectory fitting, strains can be directly computed from the voxelwise displacement data. This trajectory-based approach laid the groundwork for the development of future frameworks and toolkits, among which is the DENSEanalysis toolbox~\cite{spottiswoode2007tracking,gilliam2021denseanalysis}, a software package for cine DENSE processing widely used across research institutions.

Building on these tracking methods, recent efforts have shifted toward developing fully automated pipelines that integrate material point tracking -- a method that computes the trajectories of discrete myocardial voxels as they move throughout the cardiac cycle -- and strain computation. Gilliam and Epstein~\cite{gilliam2012automated} developed a fully automated pipeline integrating material point tracking and strain computation using radial basis function interpolation.  This framework offered a complete solution without manual input or pre-trained machine learning (ML) models. Ghadimi et al.~\cite{ghadimi2021fully} developed a fully automated 2D DENSE strain analysis pipeline consisting of CNN-based segmentation and phase unwrapping, followed by displacement and strain computation using the material point tracking method originally proposed by Spottiswoode et al.~\cite{spottiswoode2007tracking}. Following their previous work, Ghadimi et al.~\cite{ghadimi2023improved} proposed a regularized spatiotemporal least squares (RSTLS) formulation in 2D eliminating the need for explicit trajectory models (i.e., spatio-temporal polynomial interpolation or Fourier-based models), enforcing instead spatial and temporal smoothness directly on the Lagrangian displacement field. While these methods exemplify a shift from manually guided or model-constrained approaches toward automated frameworks, significant drawbacks still exist: methods are usually restricted to 2D, slice-by-slice evaluation, they are often applied and validated only on small cohorts or mid-ventricular regions with high image quality, and frequently rely on manual user intervention for the initial segmentation. Furthermore, these approaches can be computationally expensive and sensitive to the signal-to-noise ratio (SNR) nature of DENSE data.

Among the challenges identified in automating these frameworks, the segmentation of the myocardium remains a primary problem, leading to the exploration of both classical computer vision and ML techniques. While recent efforts have primarily focused on neural networks, earlier approaches applicable to DENSE MRI also explored classical computer vision techniques. In this context, Angus-Barker and Kar~\cite{angus2017mri} developed an automated boundary detection method that combined multi-level Otsu thresholding with displacement-guided temporal tracking, demonstrating good agreement with manually selected endocardial and epicardial contours. Most recently, deep learning (DL) has revolutionized medical image segmentation, and MR image processing has particularly benefited. To improve segmentation performance in cine DENSE and other cardiac MR images, researchers have pursued new ML model architectures and the use of richer input data beyond standard magnitude images. Barbaroux et al.~\cite{barbaroux2023automated} have developed a spatiotemporal convolutional neural network for automated segmentation of both long- and short-axis DENSE MRI, which takes advantage of segmentation coherence through time. Their results showed the benefits of spatiotemporal modeling over purely 2D CNNs, which often struggle with poor contrast in certain phases like early systole and end-diastole. In addition to changes in model architecture, several studies have enhanced segmentation by incorporating additional imaging data beyond magnitude images (e.g.,~\cite{shen2021automated,wu2021fast}). Specific to DENSE MRI segmentation, in our previous work~\cite{jahromi2024nnu}, we also demonstrated that combining magnitude and phase channels can enhance segmentation accuracy by providing additional spatial cues and leads to improved delineation of the left ventricular myocardium and cavity throughout the cardiac cycle. This strategy is particularly effective in challenging regions such as the basal and apical slices or during early systole.

Once an acceptable segmentation of the cine DENSE data is available, the next step is to estimate the Lagrangian displacement and strain fields. While DENSE data is initially time-resolved and acquired as a series of 2D slices (2D + time), stacking the short-axis slices as a 3D volumetric representation (3D + time) increases the accuracy of strain calculations and provides a direct measure of longitudinal strains.

In this work, we extend and combine previous approaches in ML, displacement processing, and modeling to design and validate a fully automated 3D pipeline for myocardial strain estimation from DENSE MRI. Our goal is to improve some of the drawbacks (i.e., lack of full automation, limit to 2D applications, lack of full ventricular coverage) still present in strain calculation from DENSE MRI. Our workflow integrates a DL–based segmentation with phase and magnitude inputs~\cite{jahromi2024nnu}, an extension of the RSTLS method proposed by~\cite{ghadimi2023improved}, a customized phase unwrapping, and full‐stack (base‐to-apex) strain mapping (Section~\ref{sec:Method}). The proposed framework is then validated using a computational phantom and is applied to a set of forty-seven (N=47) healthy volunteers stratified by age and sex, reporting circumferential, radial, and longitudinal strain across the entire left ventricle (Section~\ref{sec:Results}).


\section{Method}
\label{sec:Method}

\begin{figure*}[t]
  \centering
  \includegraphics[width=1\linewidth]{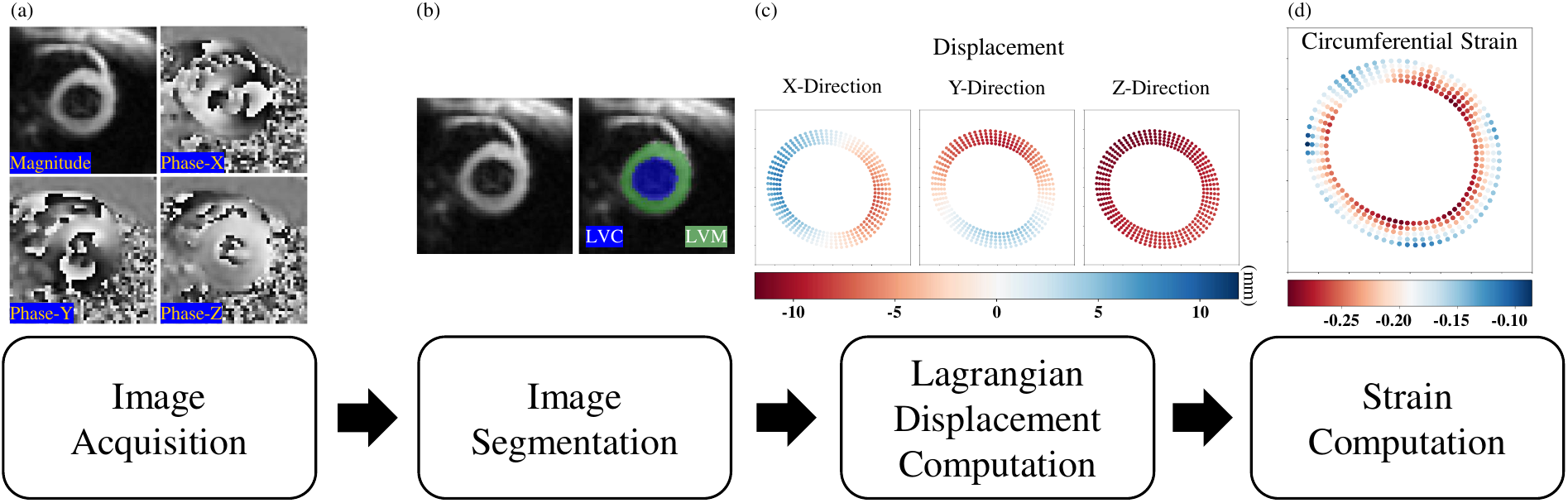}
  \caption{Overview of the proposed fully automated 3D DENSE analysis pipeline. 
  (a) \textbf{Image Acquisition}: Representative input data consisting of a magnitude image and phase-encoded displacements in three directions (Phase-X, Phase-Y, and Phase-Z).
  (b) \textbf{Image Segmentation}: Multi-class segmentation of the left ventricle, where the left ventricular myocardium (LVM) and left ventricular cavity (LVC) are segmented using the nnU-Net framework.
  (c) \textbf{Lagrangian Displacement Computation}: Constructing displacement fields in the X, Y, and Z directions (in mm).
  (d) \textbf{Strain Computation}: Lagrangian strain maps (demonstrated here with a peak-systolic circumferential strain map).}
  \label{fig:pipeline}
\end{figure*}

The overall methodology consists of four steps: 1) Image acquisition; 2) Image Segmentation; 3) Lagrangian Displacement Computation; and 4) Strain Computation (Fig.~\ref{fig:pipeline}). Each step is described in detail in Sections~\ref{sec:ImageAcquistion} to~\ref{sec:StrainCalc}, together with the validation of the proposed framework in Section~\ref{sec:validation} and the statistical analysis method in Section~\ref{sec:statistical}.

\subsection{Image acquisition and preprocessing}
\label{sec:ImageAcquistion}

The dataset used in this study consists of DENSE images acquired from a cohort of healthy volunteers (N=47) that was prospectively obtained under Institutional Review Board (IRB) oversight with signed statements of informed consent from each subject (see the \hyperlink{sec:ethics}{Ethics Statement}).
The imaging data used in this study are part of the publicly available repository at \url{https://purl.stanford.edu/xt223hz2438}.
This dataset is partitioned into two independent cohorts:

\begin{itemize}
    \item \textbf{A training and validation dataset}, used for developing the segmentation model. This dataset consists of 20 volunteers, partitioned into 80\% for training and 20\% for validation within the cross-validation (i.e., 5-fold) framework, comprising 130 short-axis slices and totaling 3789 cardiac frames.
    \item \textbf{A hold-out inference-only dataset}, used exclusively for inference with the trained segmentation model. This set consists of 27 volunteers, comprising 216 short-axis slices and totaling 5764 cardiac frames.
\end{itemize}

The demographic distribution of the full cohort analyzed in this study ($N=47$) is summarized in Table~\ref{tab:demographics}.

\begin{table}[t]
\centering
\small
\setlength{\tabcolsep}{4pt}
\caption{Demographic breakdown of the healthy volunteers' study cohort. Race and ethnicity were self-reported at enrollment.}
\label{tab:demographics}
\begin{tabular*}{\columnwidth}{@{}l@{\extracolsep{\fill}}ccc@{}}
\toprule
Characteristic & $<$40 years & $\geq$40 years & Total \\
\midrule
\textit{Sex} & & & \\
\quad Male   & 12 & 11 & 23 \\
\quad Female & 12 & 12 & 24 \\
\addlinespace
\textit{Race} & & & \\
\quad Asian                            & 9 & 13 & 22 \\
\quad Black/African-American           & 1 & 0 & 1 \\
\quad Mixed Race                       & 4 & 3 & 7 \\
\quad White                            & 10 & 5 & 15 \\
\quad \makecell[l]{Prefer not to report/\\Unknown} & 0 & 2 & 2 \\
\addlinespace
\textit{Ethnicity} & & & \\
\quad Hispanic or Latino               & 4 & 3 & 7 \\
\quad Not Hispanic or Latino           & 20 & 18 & 38 \\
\quad \makecell[l]{Prefer not to report/\\Unknown} & 0 & 2 & 2 \\
\midrule
Total & 24 & 23 & 47 \\
\bottomrule
\end{tabular*}
\end{table}

The DENSE image acquisition used the following parameters: all cardiac DENSE scans were ECG-gated and acquired under breath-hold conditions (breath-hold duration between 12--20 s). Images were primarily acquired with an in-plane resolution of $2.5 \times 2.5$ mm$^2$. In a small subset of the cohort (5 out of 47 volunteers), images were acquired with a higher in-plane resolution of $2.0 \times 2.0$ mm$^2$. Slice thickness is 8 mm for all subjects.

Key pulse sequence parameters included: echo time (TE) = 1.08~ms, repetition time (TR) = 15~ms with view sharing and 30~ms without view sharing, and a flip angle of $15^{\circ}$. Displacement encoding was applied in the x, y, and z directions using a simple 4-point encoding strategy with an in-plane displacement encoding frequency ($k_e$) of 0.10 cycles/mm and an out-of-plane displacement encoding frequency of 0.08~cycles/mm. 

Depending on the acquisition resolution, $2.5$ mm vs. $2.0$ mm, the resulting reconstructed image size was either $80\times80$ or $100\times100$, respectively. These resulting image resolutions are preserved throughout all stages of the study. Preserving image resolution eliminates the need for interpolation of the phase data, thereby reducing the risk of introducing distortions in regions affected by phase wrapping.

Ground-truth annotations were generated using a custom graphical user interface -- DENSE manual Segmentator -- developed specifically for extracting the myocardial region in DENSE MR images (the code for the manual DENSE segmentation is publicly available at \url{https://github.com/CBL-UCF/DENSEmanualSegmentator}). The annotation process involved manual detection of the endocardial and epicardial contours throughout the cardiac cycle to construct voxel-wise masks corresponding to the left ventricle cavity (LVC) and left ventricle myocardium (LVM) classes.

\subsection{Image segmentation}
\label{sec:ImageSegmentation}

The nnU-Net~\cite{isensee2021nnu} ML model is used to segment the LVM and LVC across the entire cardiac cycle. As a self-adapting UNet-based algorithm, nnU-Net automatically configures preprocessing steps (e.g., intensity normalization and resampling) and rule-based architectural/training parameters (e.g., depth, kernel size, patch size, and batch size) according to dataset properties, while relying on fixed default settings for others (e.g., optimizer and polynomial learning rate schedule).

Following our previous work~\cite{jahromi2024nnu}, each input file was represented as a four-channel image, consisting of the DENSE magnitude and the three phase components corresponding to the X, Y, and Z displacements. This input format allows the model to leverage both anatomical -- from the magnitude image -- and displacement -- from the phase images -- information for improved segmentation. Specifically, we adopt the 3D full-resolution configuration of the nnU-Net model, where the third dimension corresponds to the temporal dimension (i.e., the cardiac phase dimension). Accordingly, each 3D input volume is constructed by stacking the time frames corresponding to a single short-axis slice, where each frame consisted of a four-channel image. This setup allows the model to capture the temporal continuity of the myocardium and cavity and segment the LVM and LVC for all cardiac frames, including the initial and final frames where the image contrast is typically low.

During training, a five-fold cross-validation strategy is adopted to improve robustness and reproducibility across subjects. For each fold, 16 subjects are used for training and four subjects are employed for validation, ensuring that the model validation occurred on completely unseen subjects in every fold. The model loss function consists of a combination of Dice similarity coefficient and cross-entropy. Training was performed for up to 1000 epochs per fold, with automatic learning-rate scheduling to facilitate convergence. This conservative number of epochs was chosen to ensure that the training was successfully completed. The best-performing model from each fold, selected based on its performance on the validation set, is used in the ensemble strategy. For final inference, predictions from all five trained models are combined using an ensemble strategy, in which the voxel-wise softmax probabilities are equally averaged across folds before generating the final segmentation. Training was conducted on a single NVIDIA A30 GPU with 24GB of memory, and on average, training each fold required approximately 5 hours. For final inference, the average computation time is $\approx$5.43 ms per frame. The subsequent displacement and strain analysis is performed on the segmentations from both the cross-validation and the hold-out datasets.

In rare cases (fewer than 20 frames out of 9,553 across all subjects) where the segmentation failed to yield valid epicardial or endocardial contours (e.g., when the myocardium contour is not closed or is missing entirely), a post-processing step based on temporal consistency is applied. Problematic frames, typically observed in the basal slices near peak systole, where the valve structure can emerge, or early systolic phases (where the image contrast is low) are corrected by reconstructing the missing contours. When invalid frames occur between valid neighboring frames, epicardial and endocardial boundaries are linearly interpolated from the adjacent contours in time based on their time frame. If the invalid frame occurs at the beginning or at the end of cardiac cycle, the nearest valid contours are used for the invalid frame. This procedure ensures anatomically plausible masks across the entire cardiac cycle and prevents local segmentation failures from affecting the subsequent displacement and strain computations.

In the following analyses, when referring to the peak systolic frame, the latter is defined by computing the frame with the smallest LVC per slice, and then taking the median of all these frames for a given subject.

\subsection{Lagrangian displacement computation}
\label{sec:DispCalc}
Following the generation of segmentation masks for all subjects in both the cross-validation and test cohorts, the next step in our pipeline consists in computing Lagrangian displacements. To this end, we extend the regularized spatiotemporal least squares (RSTLS) methodology proposed by Ghadimi et al.~\cite{ghadimi2023improved} to 3D. The resulting pipeline is capable of handling a full stack of short-axis DENSE slices, encompassing the left ventricle from base to apex, leading to a comprehensive evaluation of the left ventricle motion.

Following segmentation, phase unwrapping is applied to the myocardial phase data to resolve phase discontinuities \cite{herraez2002fast}. To fully automate this process, it is important to eliminate the need for a user-defined reference point. In our approach, the temporal consistency of the unwrapped phase is enforced by checking for frame-to-frame shifts and by applying corrections (i.e., by shifting the phase of the entire slice at the incorrectly unwrapped frame) as needed to ensure coherent displacement tracking over the cardiac cycle. After phase unwrapping, the Eulerian displacement components ($u_i$) in the X, Y, and Z directions at each voxel are computed as $u_{i} = \phi_{i} /{(k_{\text{e},i} v_{i})}$. Therein, $\phi_{i}$ is the unwrapped phase in the $i$ direction measured in \textit{cycles}, $k_{\text{e},i}$ is the encoding strength measured in \textit{cycles/mm} in direction $i$, and $v_{i}$ is the voxel size in direction $i$ measured in \textit{mm/voxel}. The resulting displacement $u_{i}$ is expressed in voxel units. Physical displacements in millimeters can be obtained by multiplying voxel displacement by the voxel size.

Lagrangian displacements are computed using the RSTLS formulation proposed by Ghadimi et al.~\cite{ghadimi2023improved} and briefly described herein with proposed modifications. For each cardiac frame, following Ghadimi et al.~\cite{ghadimi2023improved}, the Lagrangian displacements are computed at defined grid points (e.g., the voxel centers) by solving:

\begin{equation}
\begin{split}
\arg\min_{\mathbf{U}_f} \big( & \left\| \mathbf{A} \mathbf{U}_f - \mathbf{u}_f \right\|^2 + \left\| \lambda (\mathbf{B} \mathbf{U}_f) \right\|^2 \\
& + \left\| \mu ( \mathbf{U}_f - \mathbf{U}_{f-1}) \right\|^2 \big) \, ,
\end{split}
\label{eq:rstls_expanded}
\end{equation}

\noindent
where $\mathbf{U}_f$ denotes the vector containing the unknown Lagrangian displacements at cardiac frame $f$; $\mathbf{u}_f$ represents the vector containing the known Eulerian displacements obtained from the unwrapped DENSE phase at frame $f$; $\mathbf{A}$ is the spatial interpolation matrix that maps the Lagrangian displacements to the Eulerian measurements; and $\mathbf{B}$ is the matrix form of a spatial smoothing operator. The parameters $\lambda$ and $\mu$ weight the importance of spatial and temporal smoothness, respectively. For notational simplicity, $\lambda$ is written here as a scalar, although in our implementation it is direction-dependent, as subsequently discussed. The first term in Eq.~\ref{eq:rstls_expanded} aims to minimize the difference between the measured Eulerian displacements and the mapped Lagrangian displacements. 

In our implementation, we extend the operators represented by the matrices $\mathbf{A}$ and $\mathbf{B}$ to 3D and enforce spatial smoothness using evenly distant grid points to avoid a local bias in the computed Lagrangian displacement field, especially near the LVM boundaries.

In this work, we optimized the spatial and temporal regularization parameters empirically based on the image acquisition properties and performance of the pipeline on the volunteer dataset. The pipeline receives a single spatial smoothing factor, $\lambda$, which is applied to the in-plane ($x$-$y$) directions, while the weight along the $z$-axis is internally set to $\lambda/2$. This reduces longitudinal weighting accounts for the anisotropic resolution of the DENSE acquisition ($2.5 \times 2.5$ mm$^2$ in-plane versus 8 mm out-of-plane) and for the tapering of the ventricle toward the apex, mitigating undesirable smoothing of myocardial displacements against adjacent non-myocardial grid points across the larger voxel spacing. As the three strain components differ in their sensitivity to spatial regularization~\cite{ghadimi2023improved}, results of this work are reported using two sets of analyses. 
Circumferential and longitudinal strains and the Jacobian $J$ are generated using $\lambda=4$, while radial strains are computed using $\lambda=2$, since the stronger spatial smoothness induced with $\lambda=4$ suppresses the radial strain transmural gradient (see Appendix, Fig.~\ref{fig:temporal_lambda_mu_appendix}).
The temporal smoothing factor $\mu$ is set to 0.05 to ensure consistent displacement tracking throughout the cardiac cycle.

Prior to computing the 3D Lagrangian displacements, we implement an automated rigid-body translation to compensate for slice misalignment caused by factors such as inconsistent breath-hold positions. This alignment is performed using the following procedure. First, the \textit{resting frame} for each slice is constructed by mapping the epicardial and endocardial contours (from first frame up to the peak systolic frame) back to their original reference configuration using the corresponding Eulerian displacement fields. These mapped contours are then averaged to generate a single resting geometry for each slice. Second, each short-axis slice is rigidly translated in plane to align its centroid along a common longitudinal axis.
Following this alignment, the Lagrangian displacement field is computed using the RSTLS approach.
Although displacements are calculated at all grid points across the image domain, the final analysis is restricted to the myocardial resting region.
This resting region serves as a spatial mask, ensuring that strain values are only computed and reported for points located within the actual myocardial tissue in the reference configuration.

\subsection{Strain computation}
\label{sec:StrainCalc}
To calculate strains, the discrete Lagrangian displacement field computed at the center of each myocardial voxel (the chosen grid points in our application of the RSTLS method) is interpolated using a global cubic radial basis function (RBF) with an added linear polynomial term and a small smoothing factor of $0.01$, enabling continuous displacement estimation at arbitrary locations within the myocardium.  Query points are then generated along radial lines extending from the centroid of the myocardium at each acquired short-axis location.  Radial lines are placed at $5^{\circ}$ angular increments, and the segments between the endocardial and epicardial intersections on each line are divided into five equal sub-segments. A query point is then placed at the center of each sub-segment, allowing strain evaluation at endocardial, mid-wall, and epicardial subregions, corresponding, respectively, to the query point in the middle of the most endocardial, mid-wall, and most epicardial segment. At these query points, the deformation gradient tensor $\mathbf{F}$ is computed and used to derive the Green--Lagrange strain tensor $\mathbf{E} = 1/2 (\mathbf{F}^\text{T}\mathbf{F}-\mathbf{I})$.
The strain tensor is then projected along the radial, circumferential, and longitudinal directions to compute the radial $E_{\text{rr}}$, circumferential $E_{\text{cc}}$, and longitudinal $E_{\text{ll}}$ strain components. The radial and circumferential directions are computed based on a polar reference system with the origin at the centroid of the analyzed short-axis slice, while the longitudinal axis is the normal to the $x$-$y$ short axis plane. 
The deformation gradient tensor $\mathbf{F}$ is also used to compute the Jacobian $J = \det(\mathbf{F})$, which is a measure of the myocardium volumetric deformation, i.e., $J=dv/dV$, where $dv$ is the infinitesimal volume in the current configuration versus the corresponding infinitesimal volume in the reference configuration $dV$.

\subsection{Framework validation}
\label{sec:validation}

To validate the proposed framework, we used a computational phantom (CP) originally proposed in~\cite{verzhbinsky2019estimating,perotti2021estimating}. The phantom is constructed as a cylindrical volume of height 40~mm, with DENSE images synthesized at short-axis locations $z=8$~mm, $z=16$~mm, and $z=24$~mm.  

Moreover, to ensure that the phantom data is processed identically to the real datasets, the generated Eulerian displacements, including noise, are saved as DICOM images and subsequently passed through the full segmentation, unwrapping, and Lagrangian displacement computation steps. This procedure guarantees that the analytical validation mimics the processing steps used to process experimentally acquired data and estimates the end-to-end accuracy of the automated pipeline under identical circumstances.

\begin{figure*}[t]
\centering
\includegraphics[width=1\linewidth]{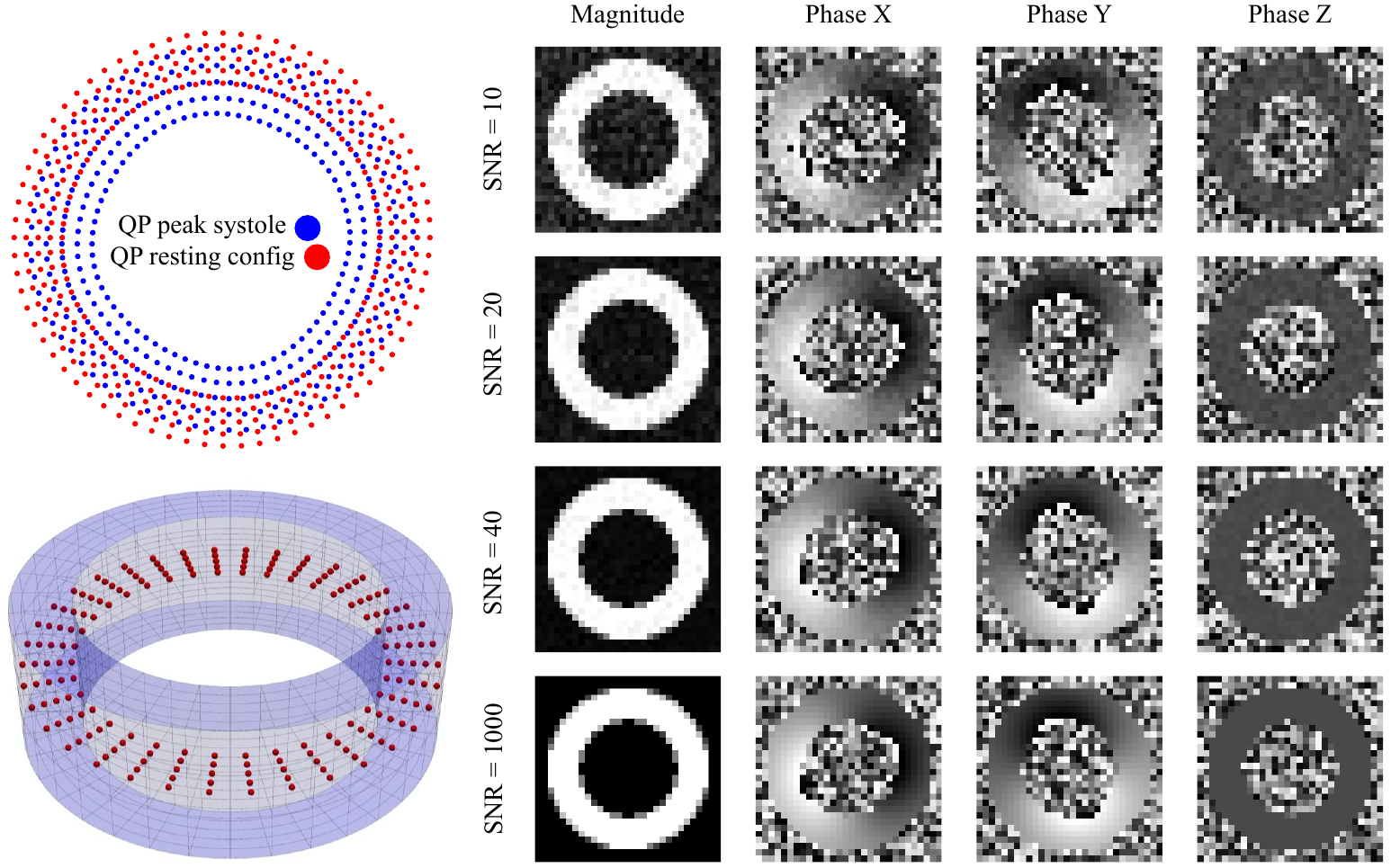}
\caption{Computational deforming phantom used to validate the framework. The cylindrical volume (height = 40~mm) is sampled at $z=8$~mm, $z=16$~mm, and $z=24$~mm to generate pseudo-experimental displacement fields.
Although the phantom spanned 40~mm in height, only 24~mm (corresponding to three voxels along the $z$-axis) are visualized here. The query points (QP) are visualized at the location of the mid slice. The phase data corresponding to the Eulerian displacements and the magnitude data corresponding to the location of the myocardium are exported as DICOM images and subsequently processed using the proposed pipeline. This allows computing the displacement and strain errors at different SNRs when the proposed pipeline is applied to analytically generated pseudo-experimental DENSE images.}
\label{fig:phantom_validation}
\end{figure*}

Four signal-to-noise ratios (SNRs) are utilized in the phantom test, i.e., SNR equal to 10, 20, 40, and 1000, with the latter corresponding to a practically noise-free signal (see Fig.~\ref{fig:phantom_validation}). Both displacements and strains in the circumferential, longitudinal, and radial directions computed using the proposed framework are compared with the known ground truth in the phantom, and the errors between computed and ground truth values are reported for the mid-slice (at $z=16$~mm).

\subsection{Statistical analysis}
\label{sec:statistical}
Statistical analysis is performed to aid the interpretation of strain values computed across the healthy volunteer cohort.
For each of the 47 volunteers, a single global median value at peak systole is computed for the circumferential and longitudinal strains over all myocardial points, while the radial median is computed over all mid-wall points only.
A two-way Analysis of Variance (ANOVA) is conducted for each strain metric to assess the influence of two independent between-subject factors: 
sex (male vs. female) and age group (age: $<40$ years 
vs. $\geq 40$ years), as well as their interaction effect.
Prior to the ANOVA, normality within each subgroup is verified using the Shapiro-Wilk test.
Normality is confirmed for all subgroups across all three strain metrics, justifying the use of the parametric two-way ANOVA.
All statistical analyses are performed using the Pingouin library~\cite{vallat2018pingouin} in Python 3.12, with a significance level $\alpha = 0.05$.


\section{Results and Discussion}
\label{sec:Results}

The first step of the pipeline is the accurate segmentation of the left ventricular myocardium (LVM), which defines the region of interest containing myocardial displacement information. Precise LVM detection is critical, as segmentation errors directly propagate to strain calculations.

The segmentation performance over 5-fold cross-validation is summarized in Table~\ref{tab:seg_dice} showing that the employed ML model achieves high accuracy for both the left ventricle myocardium (LVM) and cavity (LVC), with moderately higher Dice similarity coefficients for LVC. Variability across folds is low (overall Dice similarity coefficient $=0.91\pm0.01$), indicating robust generalization of the trained model.

\begin{table}[htbp]
\centering
\caption{Segmentation performance (mean $\pm$ standard deviation) over 5-fold cross-validation on the 20-subject training and validation set.}
\label{tab:seg_dice}
\begin{tabular*}{\linewidth}{@{\extracolsep{\fill}} l c c c @{}}
\toprule
Region & LVM & LVC & Overall \\
\midrule
Dice & $0.89 \pm 0.01$ & $0.93 \pm 0.01$ & $0.91 \pm 0.01$ \\
\bottomrule
\end{tabular*}
\end{table}

\subsection{Analysis of pseudo-experimental images generated using the computational phantom}
Before applying the proposed pipeline to experimentally acquired DENSE data, the framework's sensitivity to noise is analyzed using pseudo-experimental DENSE images generated with the computational phantom. 

\begin{figure*}[t]
    \centering
  \includegraphics[width=1\linewidth]{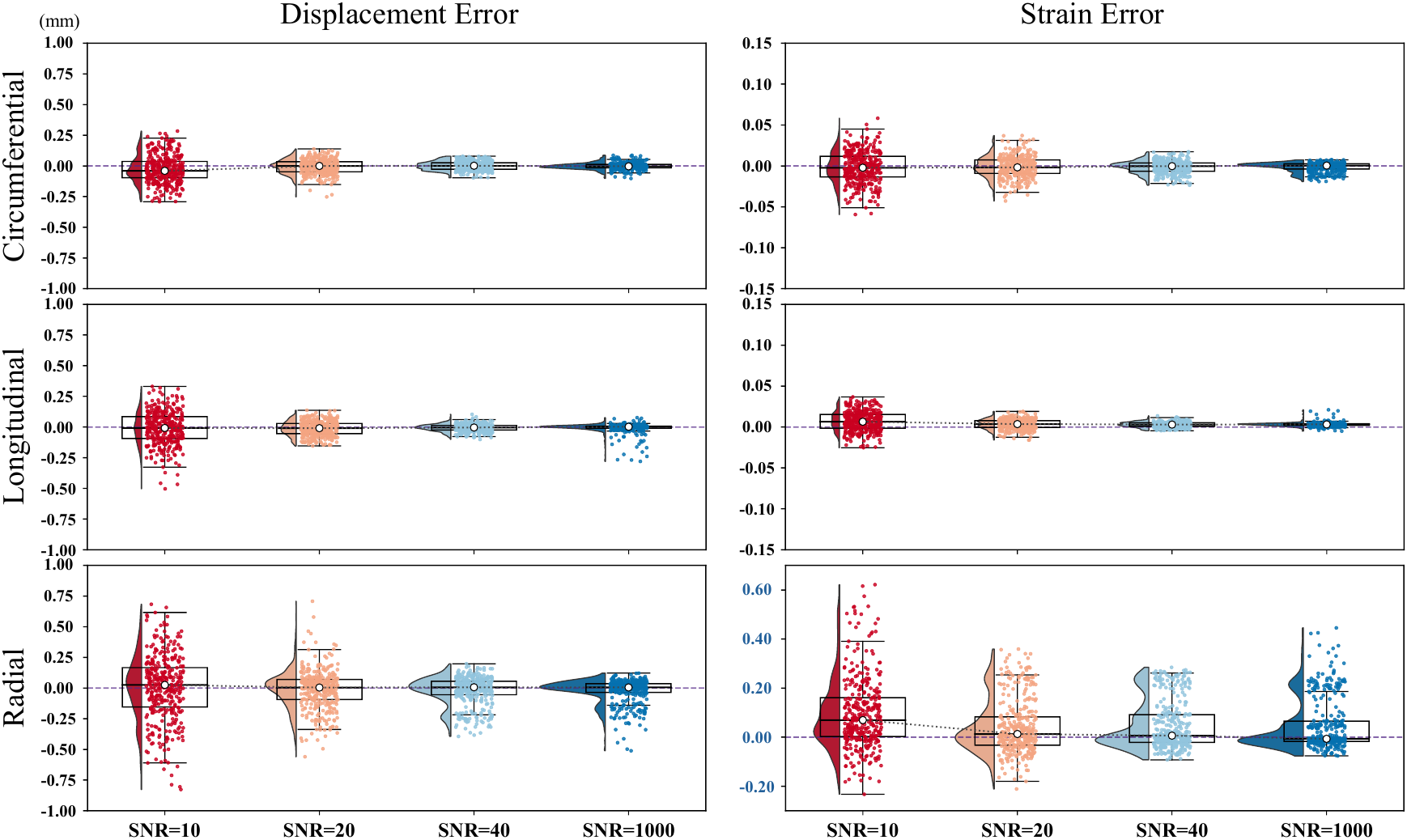}
  \caption{Effect of noise level (SNR = 10, 20, 40, 1000) on displacement and strain estimation from DENSE images generated using the computational phantom. Error distributions are visualized for the circumferential and longitudinal components, computed using $\lambda=4$, and for the radial components, computed using $\lambda=2$ (Section~\ref{sec:DispCalc}). Each plot combines the error probability density (violin plot), the inter-quartile range (box plot), and the individual voxel errors (raincloud scatter dots). Within each box plot, a white circular marker identifies the median error and a dotted line connects the medians to illustrate the trend across SNRs (a horizontal dashed line is drawn to identify zero error). All errors are computed as '(computational phantom value)$-$(computed value)'. The median displacement and median strain errors remain low and overall decrease (in absolute value) as SNR increases. Overall, circumferential and longitudinal strain errors decrease as SNR increases, while larger errors persist in radial displacements and strains even at SNR=1000, pointing to the challenges of accurately evaluating radial displacements and strains, especially close to the LVM borders (see also Fig.~\ref{fig:phantom_strain_map_snr}).}
    \label{fig:phantom_strain_disp_snr}
\end{figure*}

As shown in Fig.~\ref{fig:phantom_strain_disp_snr}, at low SNR levels (e.g., SNR = 10), the noise introduces significant errors in the computed displacement field, which then propagates to the computed strains derived from the spatial gradient of the displacements. Although all strain components are affected by noise, radial strain ($E_{\text{rr}}$) exhibits the highest sensitivity to local displacement errors. This higher sensitivity to noise may be due to the higher dependence of $E_{\text{rr}}$ on the transmural displacement gradient, where only a few voxels are present and error in segmentation at the endocardial/epicardial borders may introduce erroneous displacements or remove needed information. 

As SNR increases, a substantial reduction in displacement and strain errors is observed in both longitudinal and circumferential directions, indicating improved reliability, starting at SNR=20 and higher. However, the radial displacement and strain components continue to exhibit significant error, even at the highest SNR. This is consistent with previous observations~\cite{verzhbinsky2019estimating} where displacements and strains are computed using manual segmentation and the DENSE analysis toolbox. The persistence of strain and displacement error in the radial direction suggests that a component of these errors is not dependent on the image noise, but rather is a result of the method itself and is more likely a result of the spatial smoothing constraints within the RSTLS method. Lagrangian displacement computation in the radial direction is particularly sensitive to these constraints.

Moreover, Fig.~\ref{fig:phantom_strain_disp_snr} shows that while most median errors remain nearly constant (and approximately zero) across SNR levels, a slight decrease is observed -- most notably for the radial strain when moving from SNR = 10 to SNR = 20. In parallel, the error distribution tightens at higher SNR, highlighting improved accuracy in the lower-noise conditions. These findings underscore the importance of SNR in DENSE acquisitions and validate the robustness of the proposed computational pipeline under realistic noise scenarios for circumferential and longitudinal displacement and strain calculations.

\begin{figure*}[t]
  \centering
  \includegraphics[width=1\linewidth]{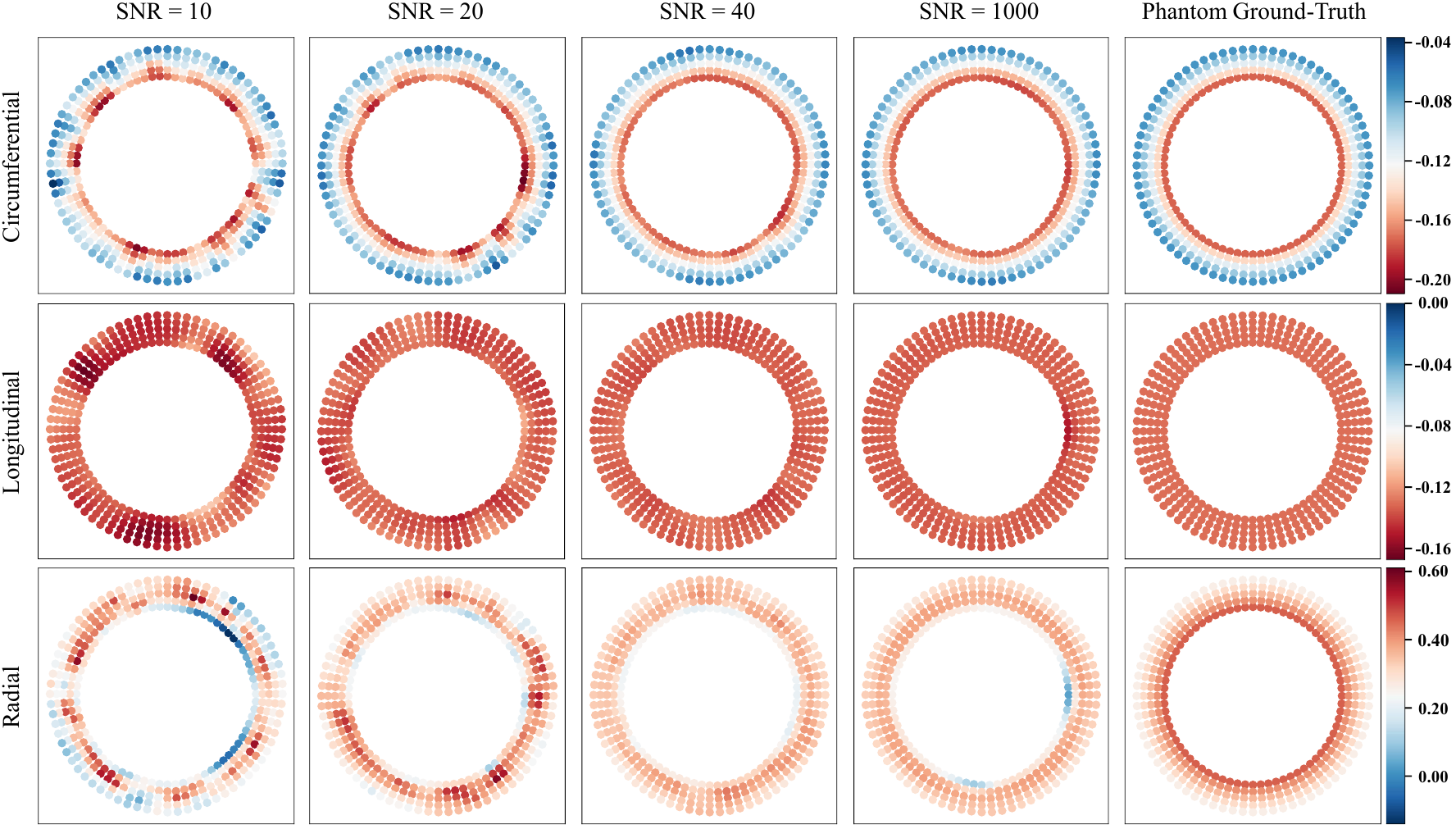}
  \caption{
  Strain maps computed by the proposed pipeline from DENSE images generated using the computational phantom across different SNR values (from left to right: SNR = 10, 20, 40, 1000). The strain maps are compared against the ground-truth strain values obtained directly from the computational phantom (right column). Circumferential (top) and longitudinal (middle) strain components approach the ground truth values as SNR increases, while the localized discrepancies at SNR=1000 for the longitudinal strains are due to misclassified voxels in the segmentation of the myocardium (see more details in Appendix~\ref{sec:appendix_A}). Larger discrepancies persist in radial strains, especially in the subendocardial layer due to spatial smoothing. In order to decrease the effect of spatial smoothing in the calculation of radial strains, the latter are computed using $\lambda=2$, while circumferential and longitudinal strains are computed using $\lambda=4$ (see Section~\ref{sec:DispCalc}).
  }
  \label{fig:phantom_strain_map_snr}
\end{figure*}

To understand if strain and displacement errors localize in a particular region (especially for the radial components), we compute the strain maps generated at peak systole from the pseudo-experimental DENSE images (Fig.~\ref{fig:phantom_strain_map_snr}). As expected, circumferential and longitudinal strains approach the ground truth values as SNR increases. Regarding radial strains, we notice that even at SNR=1000, significant strain differences persist, especially in the subendocardial layer, likely due to the spatial smoothing involving voxels with zero or small displacements proximal to the endocardial border. We also notice a localized error in both longitudinal and radial strains in the left-mid subendocardial region even at SNR = 1000. These errors are caused by a few misclassified voxels (over-estimation of the LVM mask at the endocardium) by the segmentation model that adversely affect the downstream displacement and strain computation for the idealized phantom data. A detailed analysis on the effect of removing these misclassified voxels on the final results is provided in Appendix~\ref{sec:appendix_A}.

\subsection{Application to DENSE data acquired in healthy volunteers}
\begin{figure*}[t]
  \centering
  \includegraphics[width=1\linewidth]{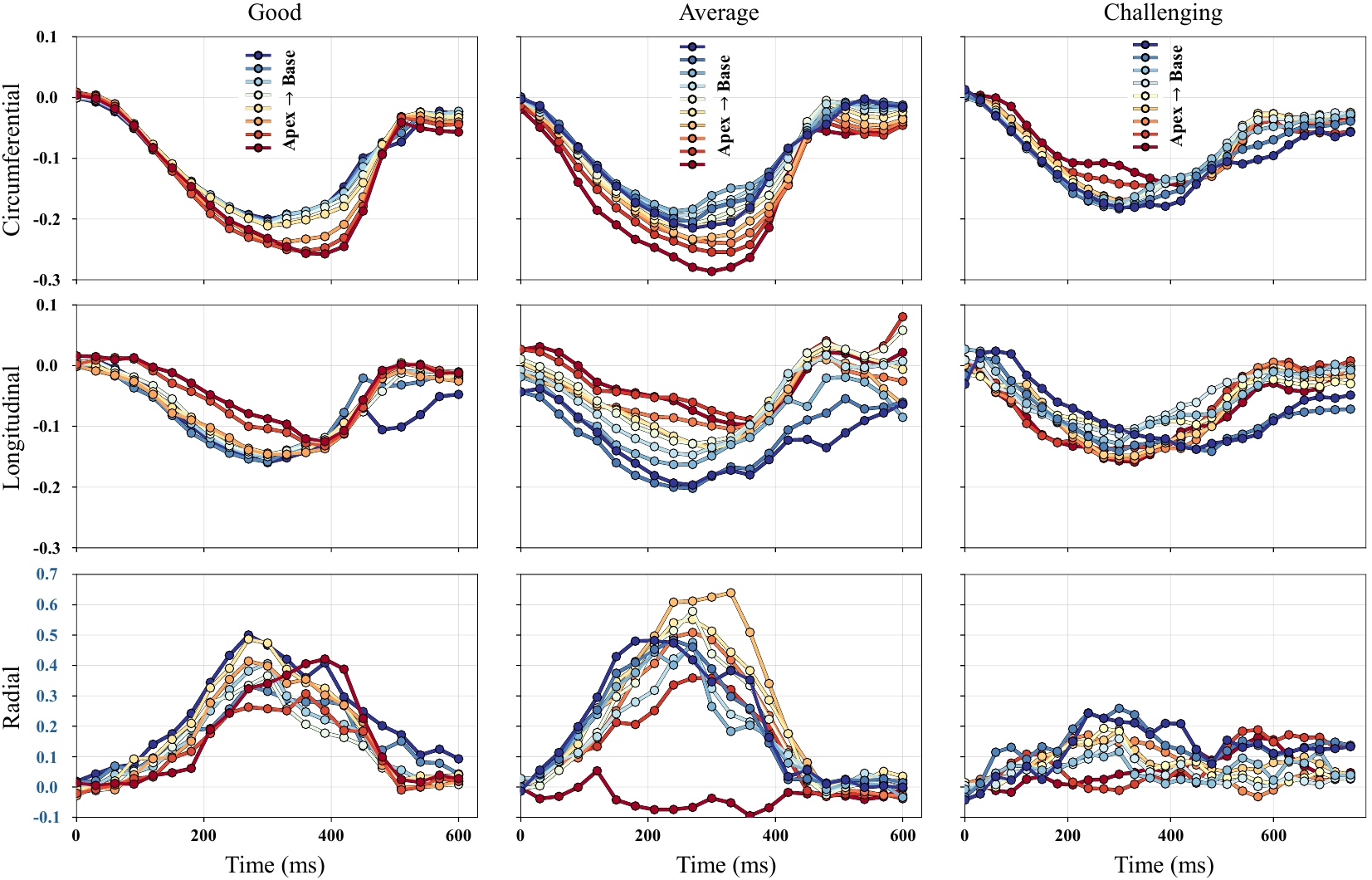}
  \caption{
  Representative cases illustrating slice-wise median strains across the cardiac cycle and from apex-to-base. Circumferential, longitudinal, and radial strains are shown for a `good', `average', and `challenging' case. Each line corresponds to the median strain computed per slice ordered from apex to base. Slice-wise circumferential and longitudinal strains are computed using $\lambda=4$ and their medians are computed across each entire slice. Radial strains are computed using $\lambda=2$ and their median is computed only across mid-wall values, where radial strains are computed more reliably.
  }
  \label{fig:strain_slice_wise}
\end{figure*}

Once the method performance is confirmed by the results of the computational phantom, the automated end-to-end pipeline is applied to compute cardiac strains from DENSE MR images acquired in healthy volunteers. Figure~\ref{fig:strain_slice_wise} shows representative (`good', `average', and `challenging') cases of slice-wise median circumferential, longitudinal, and radial strains across the cardiac cycle and for short-axis slices covering the left ventricle, from apex to base. We note a higher consistency in cardiac strains across the ventricle in the representative `good' case for all strain components. In the `average' case, the compactness of the strain curves is reduced: strains corresponding to different slices are less closely aligned, and the most apical slice shows unexpected radial strains, with the median remaining near zero or slightly negative. The challenging case demonstrates reasonable slice-wise behavior for circumferential and longitudinal strains, while the median radial strain remains low throughout the cardiac cycle, without a clear systolic peak. The poor slice-wise median radial strain in this subject is traced back to higher noise in the raw image data compared to the other two subjects.

\begin{figure*}[t]
  \centering
  \includegraphics[width=1\linewidth]{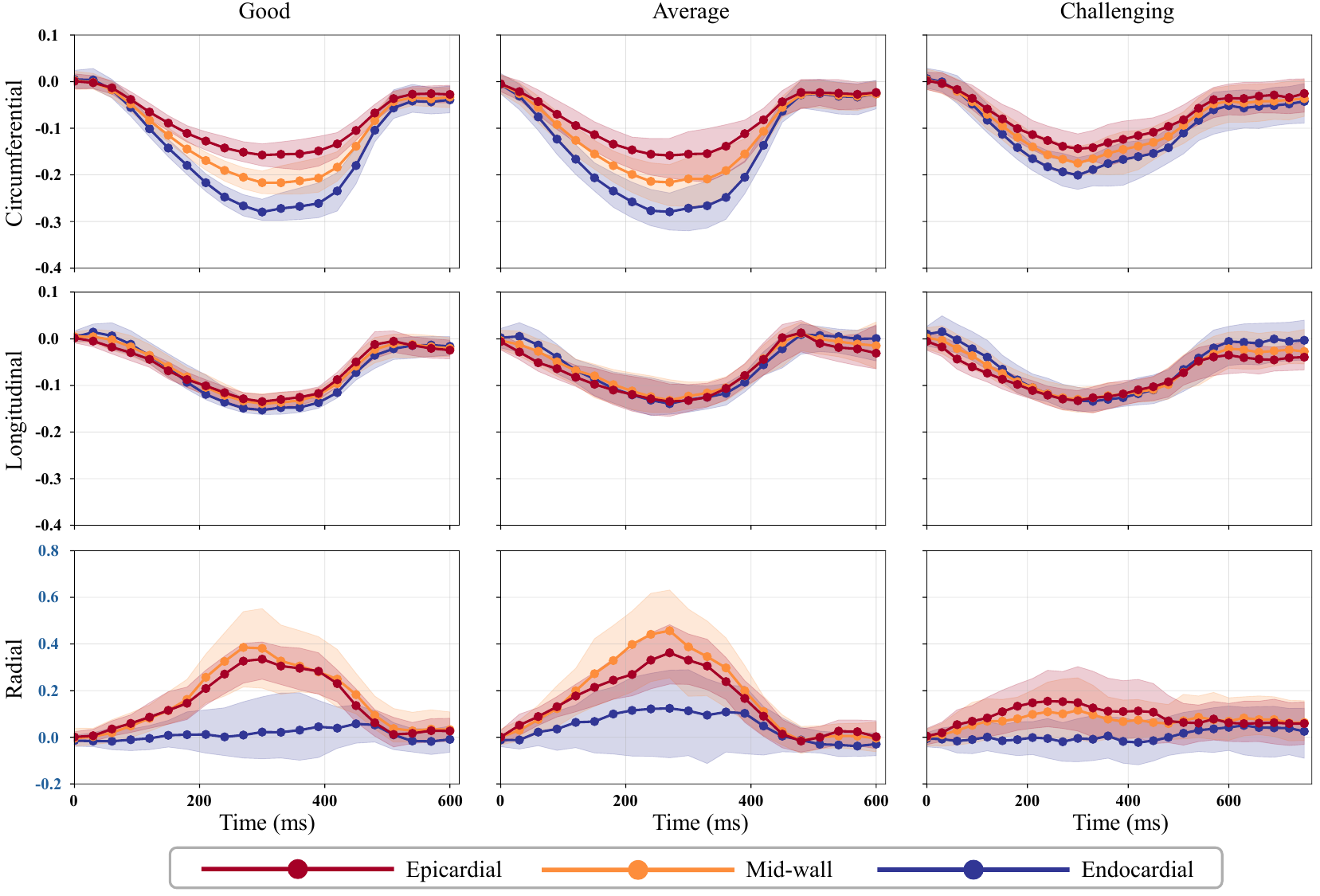}
  \caption{
  Median and interquartile range (IQR) of circumferential ($E_{\text{cc}}$), longitudinal ($E_{\text{ll}}$), and radial ($E_{\text{rr}}$) strains across the cardiac cycle for a `good', `average', and `challenging' representative subject. Each curve represents the median aggregated across all short-axis slices (from apex to base) for a specific myocardial region -- endocardial (blue), mid-wall (orange), and epicardial (red) -- with the shaded regions representing the IQR. Circumferential and longitudinal strains are computed with $\lambda=4$; radial strain with $\lambda=2$ (Section~\ref{sec:DispCalc}).
  }
  \label{fig:mid_epi_endo_wall}
\end{figure*}

For the same representative subjects, circumferential, longitudinal, and radial median strains (across all short-axis slices) are visualized in the epicardial, mid-wall, and endocardial layers (Fig.~\ref{fig:mid_epi_endo_wall}). Across all subjects, the circumferential strains exhibit the expected transmural gradient with higher values (in magnitude) in the subendocardial layer, gradually reducing toward the epicardial layer. Longitudinal strains also exhibit the expected behavior with small variations across the wall. In contrast, radial strains do not present the expected transmural gradient with higher than expected median epicardial values and near zero endocardial median values throughout the cardiac cycle. These trends confirm the results observed when processing the pseudo-experimental images obtained with the computational phantom and are likely a consequence of the spatial regularization used during Lagrangian displacement reconstruction (in the RSTLS method, displacements at a grid point are smoothed relative to all adjacent points, regardless of whether they lie within the myocardium).

\begin{figure*}[t]
  \centering
  \includegraphics[width=\linewidth]{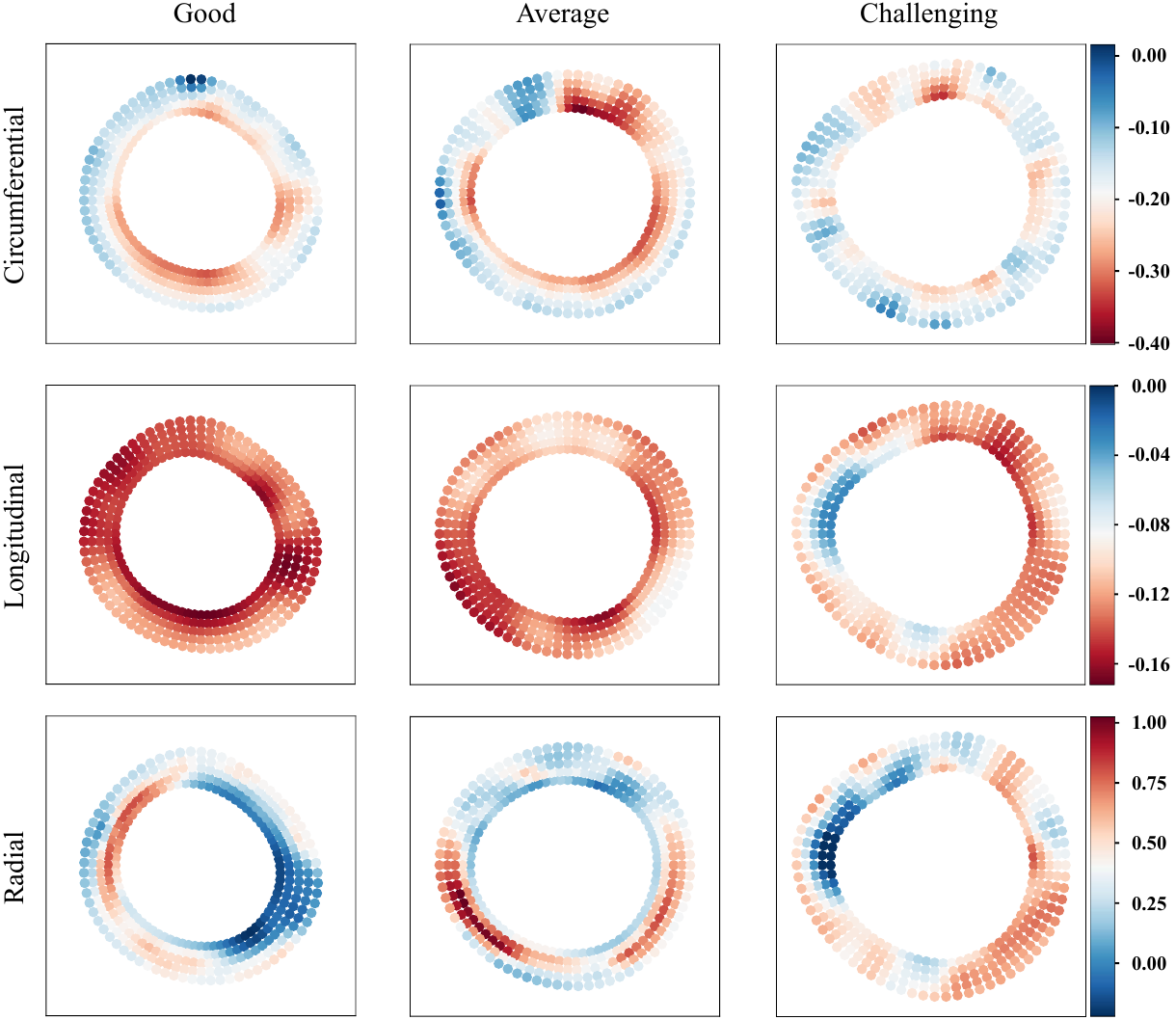}
  \caption{
  Peak systolic strain maps for a `good', `average', and `challenging' representative case. Circumferential, longitudinal, and radial strains are shown at a selected mid-ventricular slice. Circumferential strain maps present the expected transmural gradient, with larger (in magnitude) strains in the subendocardial layer. Longitudinal strain maps present small variations, although larger spatial differences emerge in the challenging case. Radial strains do not present the expected transmural gradient and are incorrectly small in the subendocardial layer due to spatial smoothing. Circumferential and longitudinal strains are computed with $\lambda=4$; radial strain with $\lambda=2$ (Section~\ref{sec:DispCalc}).
  }
  \label{fig:strain_maps_ps}
\end{figure*}

Figure~\ref{fig:strain_maps_ps} shows the peak-systolic strain maps for the representative `good', `average', and `challenging' cases.
The circumferential strain exhibits the expected pattern, with increasingly larger in magnitude (more negative) values toward the endocardial region. Longitudinal strain maps are more uniform across the wall, although progressively larger variations are present in the `average' and `challenging' cases. Radial strains do not show the expected transmural gradient and, as previously discussed, are underestimated in the subendocardial layer.

\begin{figure*}[t]
  \centering
  \includegraphics[width=1\linewidth]{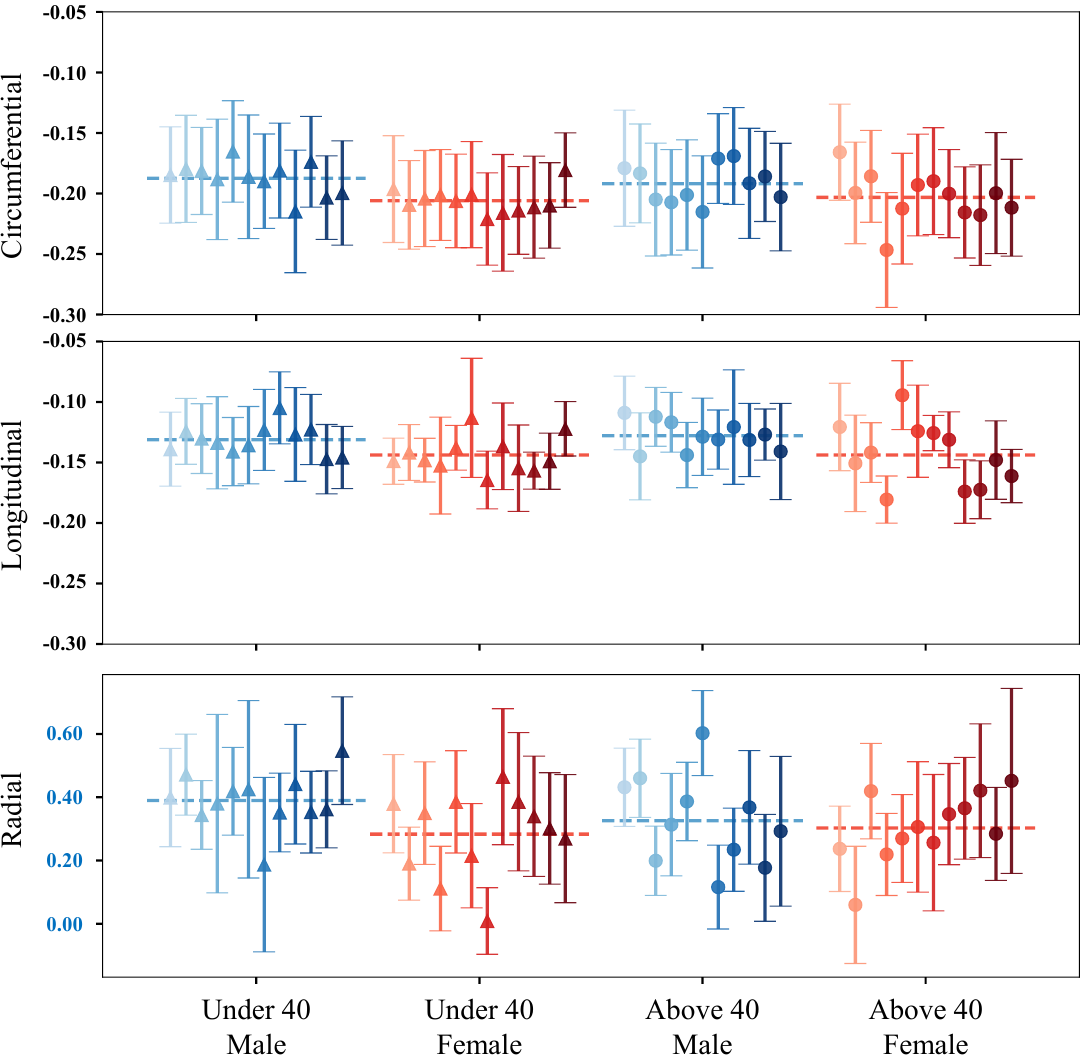}
  \caption{
  Median myocardial strain values stratified by sex and age at peak systole across all volunteers considered in this study (N=47). Circumferential ($E_{\text{cc}}$), longitudinal ($E_{\text{ll}}$), and radial ($E_{\text{rr}}$) strains are shown for four demographic groups: males under 40, females under 40, males above 40, and females above 40. Each data point represents one volunteer, with the dashed line indicating the group mean obtained as mean of all median strains in the group. Circumferential and longitudinal strains are computed with $\lambda=4$ using all myocardial points; radial strains are computed using $\lambda=2$ and median values are evaluated using only mid-wall query points.
  }
  \label{fig:strain_age_sex}
\end{figure*}

After evaluating the strain temporal and spatial distributions across representative cases, we compare strains across the forty-seven (N=47) volunteers considered in this study and stratified by sex and age (Fig.~\ref{fig:strain_age_sex}). As the previous results show that radial strains are not computed accurately across the myocardial wall, only mid-wall radial strains are used in this comparison. For each volunteer the median(IQR) strains across all slices are illustrated in Fig.~\ref{fig:strain_age_sex} and the mean of each group medians is reported in Table~\ref{tab:groups_mean_Of_medians} as an aggregate measure (the horizontal dashed lines in Fig.~\ref{fig:strain_age_sex} also represent the mean of the median strain values within each group). While the range of median circumferential and longitudinal strains is overall compact, radial strains vary more significantly across subjects, including subjects with near zero (and negative) values. The latter are unphysiological in a healthy population and confirm the challenges in accurately and consistently evaluating radial strains from DENSE MRI across a large subject cohort, where experimental noise is unavoidably present.

\begin{table*}[t]
\centering
\caption{Mean strains (computed as the mean of medians strains across volunteers) stratified by sex and age group at peak systole.
Circumferential and longitudinal strains are computed with $\lambda=4$ using all myocardial points; radial strains are computed using $\lambda=2$ and are evaluated using only mid-wall query points.}
\label{tab:groups_mean_Of_medians}
\begin{tabular}{l c c c c}
\toprule
\textbf{Strain} & \makecell{\textbf{Male} \\ ($<$ 40 y., $n=12$)}
                & \makecell{\textbf{Female} \\ ($<$ 40 y., $n=12$)}
                & \makecell{\textbf{Male} \\ ($\geq$ 40 y., $n=11$)}
                & \makecell{\textbf{Female} \\ ($\geq$ 40 y., $n=12$)} \\
\midrule
Circumferential ($E_{\text{cc}}$) & -0.187 & -0.206 & -0.192 & -0.203 \\
Longitudinal ($E_{\text{ll}}$)    & -0.131 & -0.144 & -0.128 & -0.144 \\
Radial ($E_{\text{rr}}$)          &  0.390 &  0.283 &  0.326 &  0.303 \\
\bottomrule
\end{tabular}
\end{table*}

The influence of sex and age on strain components is finally evaluated using a two-way ANOVA, with the results summarized in Table~\ref{tab:anova_results}. The analysis shows a significant effect of sex on circumferential strain ($p=0.0016$) and longitudinal strain ($p=0.0071$). In contrast, there is no significant effect of sex ($p=0.0651$), age, or their interaction on radial strains. Age did not show a significant effect on any of the three strain components, and no significant interaction effects between sex and age group are observed for any metric. These results suggest that sex, rather than age, is a significant factor in circumferential and longitudinal myocardial deformation in this healthy study cohort.

\begin{table*}[t]
\centering
\caption{
Two-way ANOVA p-values for myocardial strain components. Bold values indicate statistical significance ($p < 0.05$). Circumferential and longitudinal strains are computed with $\lambda=4$ using all myocardial points; radial strains are computed using $\lambda=2$ and are evaluated using only mid-wall query points.
}
\label{tab:anova_results}
\begin{tabular}{l c c c}
\toprule
\textbf{Strain} & \makecell{\textbf{Sex} \\ (Male vs. Female)} & \makecell{\textbf{Age} \\ ($<$ 40 vs. $\geq$ 40 y.)} & \makecell{\textbf{Interaction} \\ (Sex $\times$ Age)} \\
\midrule
Circumferential & \textbf{0.0016} & 0.8450 & 0.4112 \\
Longitudinal   & \textbf{0.0071} & 0.7429 & 0.7454 \\
Radial           & 0.0651 & 0.5378 & 0.2317 \\
\bottomrule
\end{tabular}
\end{table*}


\subsection{Limitations}

The proposed framework presents a pipeline towards fully automated 3D left ventricular strain analysis from DENSE MRI, but there are several limitations that should be mentioned and further addressed.

First, the most important limitation of this pipeline is the persistent challenge in accurately estimating radial displacements and strains. As demonstrated in the results obtained using the computational phantom and the healthy volunteers' data, radial strains exhibit higher error than circumferential and longitudinal strains, especially in the subendocardial region. The underestimation of radial strains in the subendocardial region likely originates from the spatial and temporal smoothing constraints within the current RSTLS implementation, which can adversely affect displacement gradients in the radial direction. These smoothing constraints, however, are essential to mitigate the impact of noise. Simply decreasing the spatial smoothing factor $\lambda$ to compute radial strains did not solve the problem (although it was beneficial) and other strategies should be investigated such as, for example, a spatially adaptive regularization strategy.

Second, and related to the previous aspect, the temporal and spatial smoothing factors adopted in this study were chosen based on the computational phantom study and based on the image quality of the analyzed dataset. This `optimization' process remains empirical and the selected values cannot be regarded as optimal. In addition, translating these parameter values to other DENSE image datasets (e.g., with different spatial resolution and SNR) remains challenging. An objective criterion/function to be optimized for the selection of the spatial and temporal smoothing factors will increase the robustness of this approach and facilitate its adoption.

Third, although this study considers a cohort of forty-seven healthy volunteers (a large cohort with respect to other studies based on DENSE MRI) and includes short-axis slices from apex-to-base (not only the ones in the midventricular region), the presented analyses are limited to healthy volunteers. The performance of the proposed pipeline on a dataset including subjects affected by different cardiomyopathies (e.g., myocardial infarction, hypertrophic cardiomyopathy) needs to be established. Before applying to a different cohort, the segmentation model may require retraining and the RSTLS approach may need to be calibrated.

Finally, while the nnU-Net model achieved good overall accuracy, misclassified voxels may still lead to significant errors in displacement and consequently strain calculations (as seen, for example, from the results obtained using the computational phantom and the images generated with SNR=1000). Improving the robustness of the segmentation model is therefore worthy of investigation.


\section{Conclusions}

We presented a fully automated framework for 3D left ventricular strain analysis from cine DENSE MRI that integrates DL-based segmentation with the regularized spatiotemporal least squares (RSTLS) method~\cite{ghadimi2023improved} for displacement computation extended to 3D. Validated on a computational phantom and applied to a cohort of 47 healthy volunteers, the pipeline provides robust circumferential and longitudinal strain measurements. The end-to-end automation of this complicated workflow provides a key step toward the broader use of DENSE MRI in clinical and research settings.


\section*{Acknowledgements}

This material is based upon work supported by the National Science Foundation under Award No. 2205043 to L.E.P. and D.W., and 2205103 to D.B.E.


\section*{Data and code availability}
The data used for this study are available as part of the ``Human MRI Data to Quantify Cardiac Performance'' repository at \url{https://purl.stanford.edu/xt223hz2438}.
The main codes used in this study are publicly available on GitHub: the full analysis pipeline is provided at \url{https://github.com/CBL-UCF/LVASD} and the manual DENSE segmentation tool is accessible at \url{https://github.com/CBL-UCF/DENSEmanualSegmentator}.


\section*{Ethics statement}
\hypertarget{sec:ethics}{}
DENSE MRI data was acquired at Stanford University under a protocol approved by the Stanford Administrative Panel on Human Subjects in Medical Research (reference numbers FWA00000929 and FWA00000934), and written informed consent was obtained from every participant prior to imaging (IRB protocol 67727, last approved on January 8, 2026). The corresponding anonymized imaging data was analyzed at the University of Central Florida under IRB protocol STUDY00004491 (last approved on May 11, 2026).


\section*{Competing interests}
The authors declare that they have no known competing financial interests or personal relationships that could have appeared to influence the work reported in this paper.

\section*{Authors' contribution}

\textbf{Mohammad Naqizadeh Jahromi:} Conceptualization, Formal analysis, Methodology, Software, Validation, Visualization, Writing -- Original Draft, Writing -- Review \& Editing.
\textbf{Ariel J. Hannum:} Data curation, Writing -- Review \& Editing.
\textbf{Zhan-Qiu Liu:} Data curation, Writing -- Review \& Editing.
\textbf{Daniel B. Ennis:} Formal analysis, Funding acquisition, Project administration, Resources, Supervision, Writing -- Review \& Editing.
\textbf{Dazhong Wu:} Conceptualization, Methodology, Project administration, Resources, Supervision, Writing -- Review \& Editing.
\textbf{Luigi E. Perotti:} Conceptualization, Formal analysis, Funding acquisition, Methodology, Project administration, Resources, Supervision, Writing -- Review \& Editing.


\appendix

\section{Supplementary material}
\label{sec:appendix_A}

\setcounter{figure}{0} 
\renewcommand{\thefigure}{A.\arabic{figure}}
\renewcommand{\theHfigure}{A.\arabic{figure}}

\begin{figure*}[t]
  \centering
  \includegraphics[width=1\linewidth]{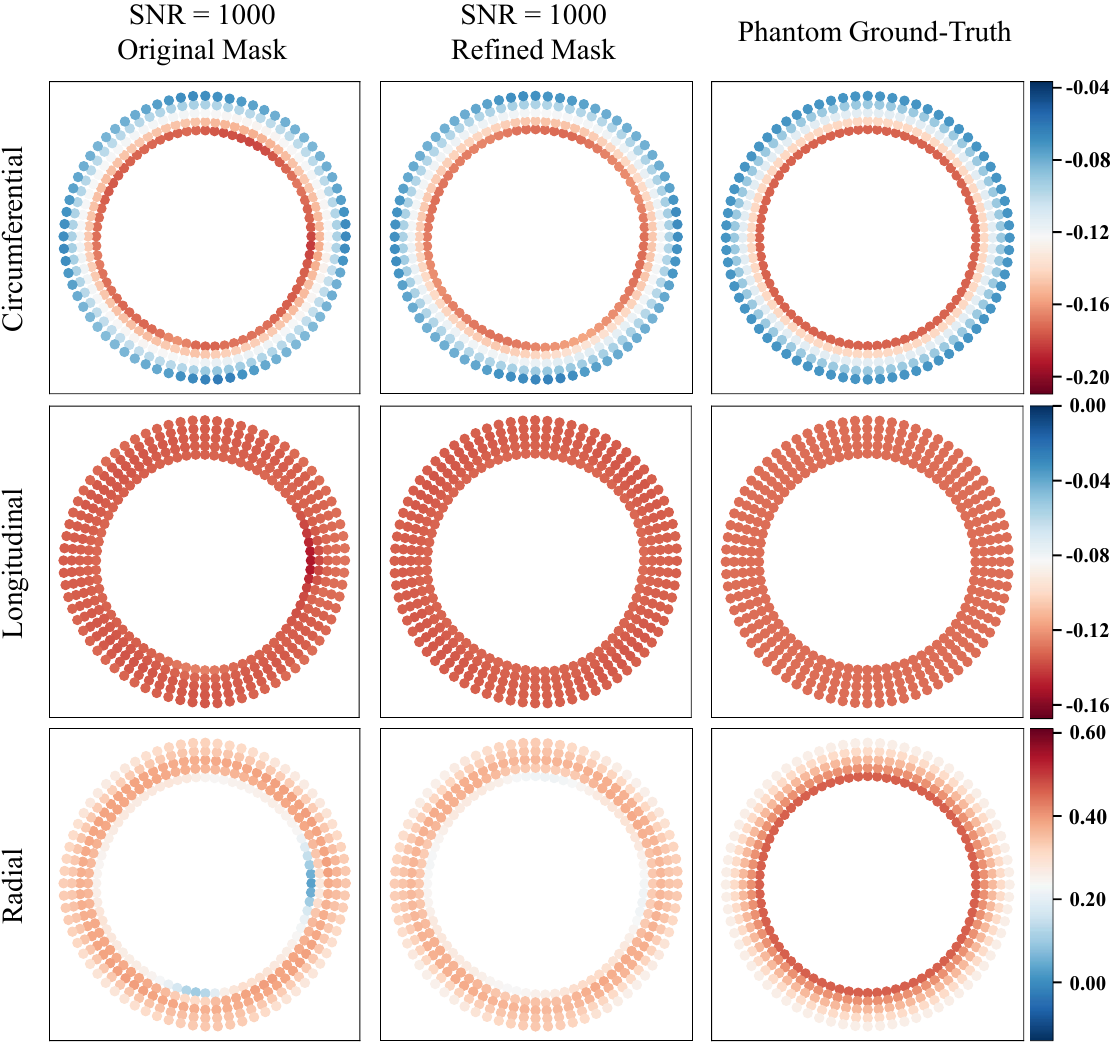}
  \caption{
  Effect of predicted mask refinement on the computed strain maps from pseudo-experimental DENSE images generated with SNR = 1000, compared against the ground-truth strains obtained directly from the computational phantom. 
  From top to bottom, circumferential, longitudinal, and radial strain components are displayed.
  Circumferential and longitudinal strains are computed with $\lambda=4$, while radial strain is computed using $\lambda=2$ (Section~\ref{sec:DispCalc}).
  }
  \label{fig:phantom_strain_mask_refine}
\end{figure*}

As illustrated in Fig.~\ref{fig:phantom_strain_mask_refine}, the manual refinement of the segmentation mask at SNR=1000, where a few misclassified voxels (four voxels in total across all three slices) are removed, has a considerable positive impact on the computed strain maps. After this correction, the localized problematic regions with large strain errors are no longer observed in the longitudinal and radial strain maps. This confirms that segmentation inaccuracies are the source of the artifacts observed in the original automated results and supports that radial strains may be especially sensitive to segmentation inaccuracies. 

\begin{figure*}[t]
  \centering
  \includegraphics[width=1\linewidth]{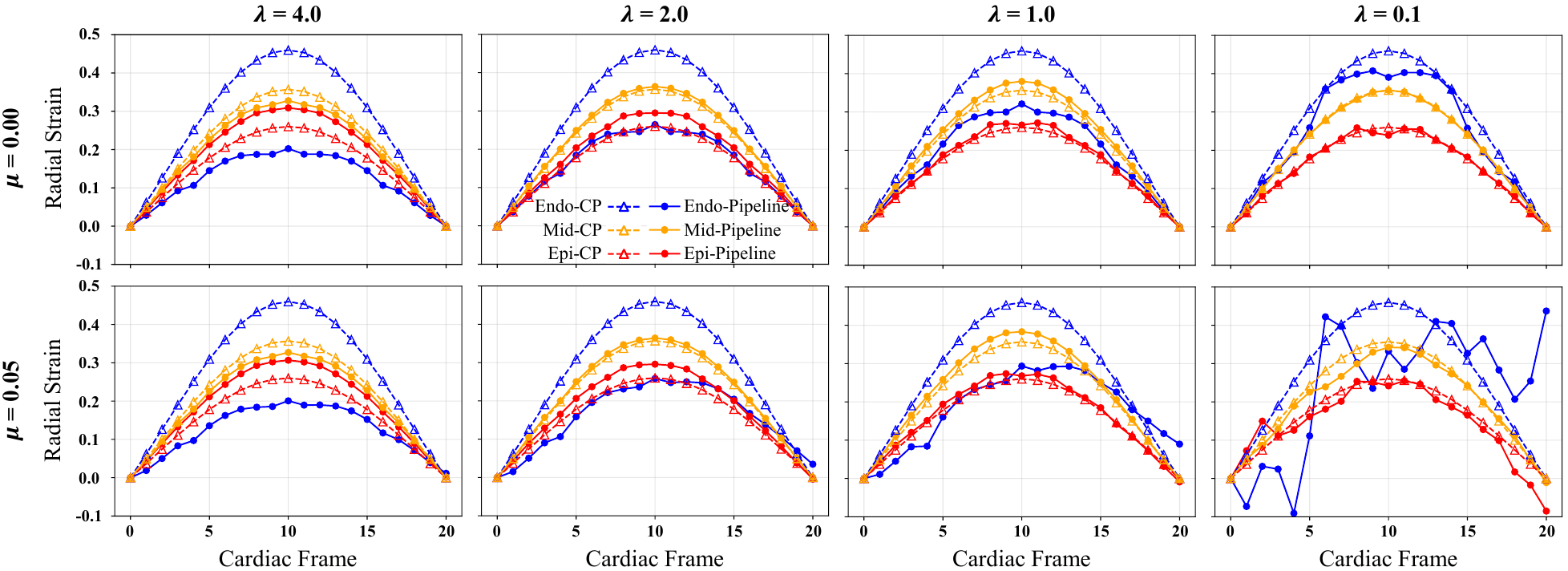}
  \caption{
    Temporal evolution of radial strains obtained using the images generated from the computational phantom with SNR=1000 and across different spatial ($\lambda$) and temporal ($\mu$) smoothing factors. The plots compare pipeline-computed radial strains against ground-truth values derived from the computational phantom (CP) for the endocardial (endo), epicardial (epi), and mid-wall (mid) layers. $\lambda$ decreases from left to right ($\lambda = 4.0, 2.0, 1.0, 0.1$), while $\mu$ is equal to $0.00$ (top row) and $0.05$ (bottom row). Larger $\lambda$ values (stronger spatial regularization) result in spatially overly smoothed displacements leading to low/incorrect transmural strain gradients. Higher temporal regularization ($\mu = 0.05$) introduces distortion with respect to the ground-truth analytical values, especially after the peak systolic cardiac phase.
  }
  \label{fig:temporal_lambda_mu_appendix}
\end{figure*}

Figure~\ref{fig:temporal_lambda_mu_appendix} demonstrates the temporal behavior of transmural median radial strains as a function of different spatial ($\lambda$) and temporal ($\mu$) smoothing factors when the proposed framework is applied to phantom data generated with SNR=1000. Smaller smoothing factors result in less deviation from the ground-truth. For the human subject data, however, the presence of larger smoothing factors is necessary to mitigate the effect of experimental noise.

\begin{figure*}[t]
  \centering
  \includegraphics[width=1\linewidth]{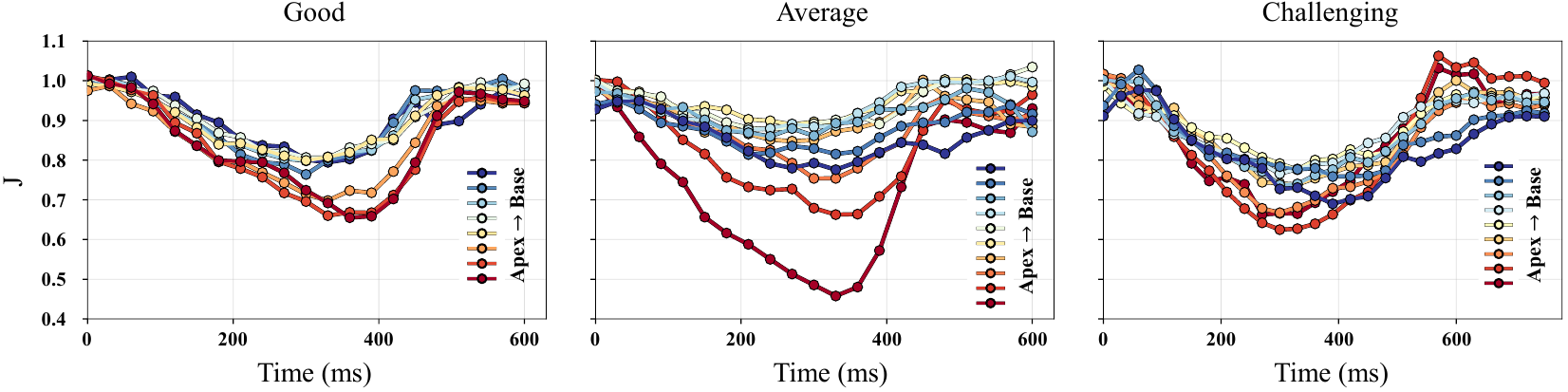}
  \caption{Representative cases illustrating slice-wise median $J$ (Jacobian of deformation gradient tensor) across the cardiac cycle. Each line corresponds to a slice, ordered from apex to base. $J$ is computed using $\lambda=4$ and reported as median of mid-wall values per slice (Section~\ref{sec:DispCalc}).}
  \label{fig:slicewise_median_appendix}
\end{figure*}

\begin{figure*}[t]
  \centering
  \includegraphics[width=1\linewidth]{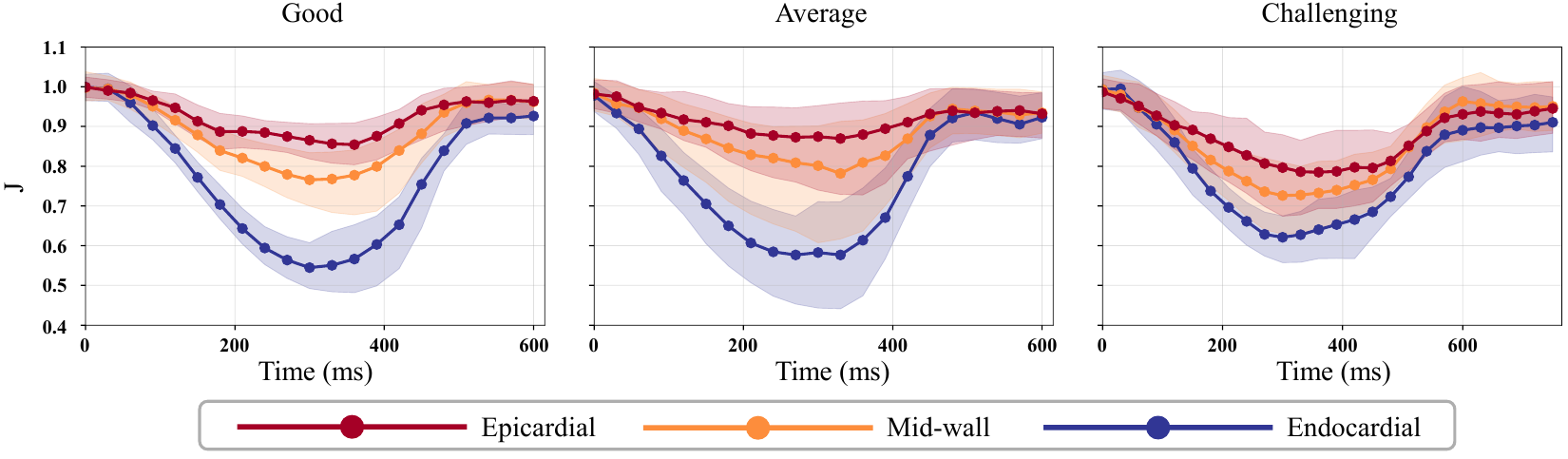}
  \caption{
  Median and interquartile range (IQR) of $J$ (Jacobian of the deformation gradient tensor) across the cardiac cycle for the three representative subjects. Each curve represents the median aggregated across all short-axis slices (from apex to base) for a specific myocardial region -- endocardial (blue), mid-wall (orange), and epicardial (red) -- with the shaded regions representing the corresponding IQR. Jacobian $J$ is computed using $\lambda=4$ (Section~\ref{sec:DispCalc}).
  }
  \label{fig:regionwise_median_appendix}
\end{figure*}

\begin{figure*}[t]
  \centering
  \includegraphics[width=1\linewidth]{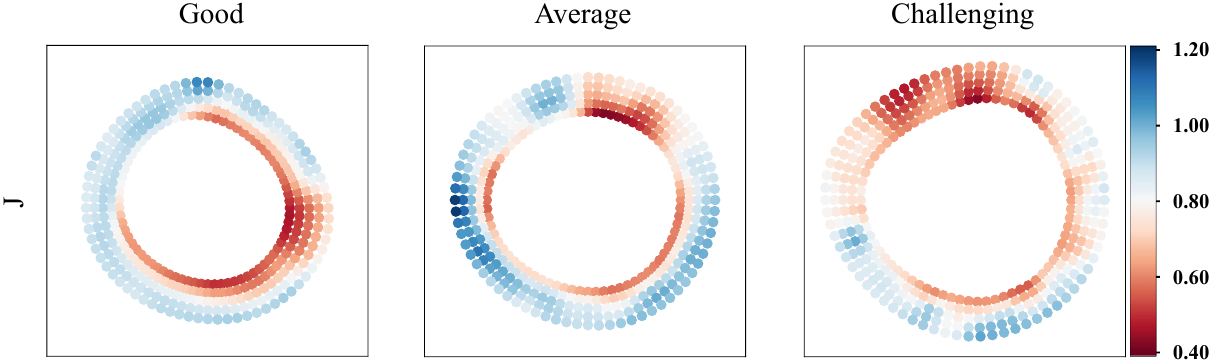}
  \caption{
  Peak systolic $J$ (Jacobian of the deformation gradient tensor) maps for representative cases at a selected mid-ventricular slice.
  Jacobian $J$ is computed with $\lambda=4$ (Section~\ref{sec:DispCalc}).
  }
  \label{fig:strain_map_appendix}
\end{figure*}

\begin{figure*}[t]
  \centering
  \includegraphics[width=1\linewidth]{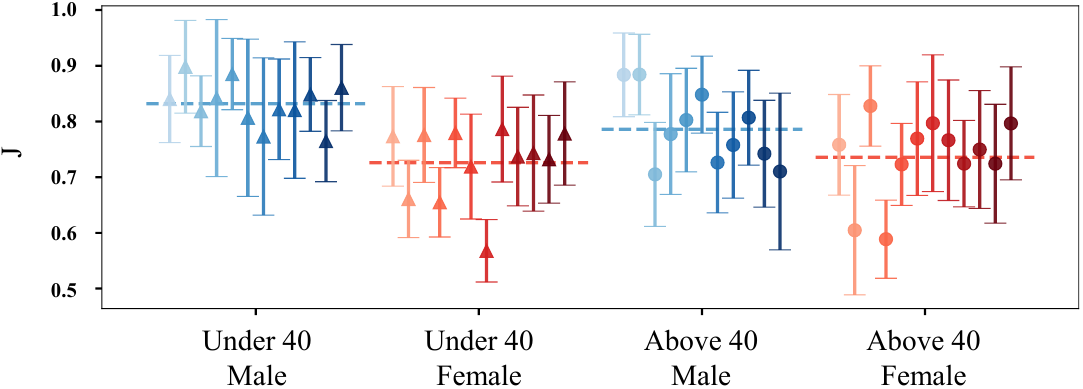}
  \caption{
  Median and mean (computed as mean of all subjects' medians within one group) myocardial $J$ (Jacobian of the deformation gradient tensor) stratified by sex and age: males under 40, females under 40, males above 40, and females above 40. Each data point represents one volunteer, with the dashed line indicating the group mean. Jacobian $J$ is computed with $\lambda=4$ (Section~\ref{sec:DispCalc}) and visualized for mid-wall points only. The mean of medians for each group is: males under 40 = 0.832, females under 40 = 0.726, males above 40 = 0.786, and females above 40 = 0.736.
  }
  \label{fig:sex_age_categorized_appendix}
\end{figure*}

Figures~\ref{fig:slicewise_median_appendix}--\ref{fig:strain_map_appendix} present the Jacobian of the deformation gradient tensor ($J$), a measure of local myocardial volume change, for the same representative subjects (`good', `average', and `challenging') analyzed in the main text. Specifically, Fig.~\ref{fig:slicewise_median_appendix} illustrates the slice-wise median mid-wall $J$ across the cardiac cycle, Fig.~\ref{fig:regionwise_median_appendix} shows the aggregated region-wise (endocardial, mid-wall, and epicardial) values, Fig.~\ref{fig:strain_map_appendix} displays the peak-systolic maps, and Fig.~\ref{fig:sex_age_categorized_appendix} provides the group-wise distribution of mid-wall $J$ stratified by sex and age.
Due to the small compressibility of the myocardium (see, e.g.,~\cite{rodriguez2006noninvasive}), we expect values of $J$ near 1. However, significant deviations from 1 in the computed values (especially at peak systole) show that $J$ is more sensitive to errors in image segmentation and displacement calculations due, for example, to experimental noise. Larger errors in $J$ tend to localize especially in the subendocardial layer and in apical slices where image segmentation and robust displacement calculations are more challenging.

\clearpage 

\clearpage


\bibliographystyle{unsrtnat}
\bibliography{references}

\end{document}